\documentclass[aps,prl,reprint,superscriptaddress,twocolumn,amsmath,amssymb,floatfix]{revtex4-2}  % TWO COLUMNS AS CLOSE TO THE PUBLISHED FORMAT AS POSSIBLE
\usepackage{soul}
\usepackage[inline,ignoremode]{trackchanges}
\tcregister{\ref}{7}
\tcregister{\pageref}{7}
\tcregister{\cite}{7}
\tcregister{\citet}{7}
\tcregister{\citep}{7}
\tcregister{\cref}{7}
\tcregister{\Cref}{7}
\soulregister\cite7
\usepackage{nth}
\usepackage{graphicx}
\usepackage{dcolumn}
\usepackage{bm}
\usepackage[left,mathlines]{lineno}
\usepackage{hyperref}
\usepackage[capitalize]{cleveref}
\usepackage{orcidlink} 
\usepackage{siunitx}
\usepackage{physics}
\usepackage[dvipsnames]{xcolor}

\begin{document}

\title{Statistical Signatures of Velocity-space Turbulence in the Earth's Magnetosheath}

\author{Tien Vo\orcidlink{0000-0002-8335-1441}}
\email{tien.vo@epss.ucla.edu}
\affiliation{Department of Earth Planetary \& Space Sciences, University of California, Los Angeles, CA 90095, USA}
\affiliation{Laboratory for Atmospheric and Space Physics, University of Colorado, Boulder, CO 80303, USA}

\author{Robert Ergun\orcidlink{0000-0002-3096-8579}}
\affiliation{Laboratory for Atmospheric and Space Physics, University of Colorado, Boulder, CO 80303, USA}
\affiliation{Department of Astrophysical \& Planetary Sciences, University of Colorado, Boulder, CO 80309, USA}

\author{Alexandros Chasapis\orcidlink{0000-0001-8478-5797}}
\affiliation{Laboratory for Atmospheric and Space Physics, University of Colorado, Boulder, CO 80303, USA}

\author{Oreste Pezzi\orcidlink{0000-0002-7638-1706}}
\email{oreste.pezzi@cnr.it}
\affiliation{Consiglio Nazionale delle Ricerche, Istituto per la Scienza e la Tecnologia dei Plasmi (ISTP), I-70126 Bari, Italy}

\author{Andrea Larosa\orcidlink{0000-0002-7653-9147}}
\affiliation{Consiglio Nazionale delle Ricerche, Istituto per la Scienza e la Tecnologia dei Plasmi (ISTP), I-70126 Bari, Italy}

\author{Sergio Servidio\orcidlink{0000-0001-8184-2151}}
\affiliation{Dipartimento di Fisica, Università della Calabria, I-87036 Cosenza, Italy}

\author{Luca Sorriso-Valvo\orcidlink{0000-0002-5981-7758}}
\affiliation{Consiglio Nazionale delle Ricerche, Istituto per la Scienza e la Tecnologia dei Plasmi (ISTP), I-70126 Bari, Italy}
\affiliation{Department of Electromagnetics and Plasma Physics, School of Electrical Engineering and Computer Science, KTH Royal Institute of Technology, Stockholm, Sweden}

\author{Le Viet Anh Doan\orcidlink{0009-0004-1115-2410}}
\affiliation{Department of Electromagnetics and Plasma Physics, School of Electrical Engineering and Computer Science, KTH Royal Institute of Technology, Stockholm, Sweden}
\affiliation{CPHT, CNRS, École polytechnique, Institut Polytechnique de Paris, 91120 Palaiseau, France}

\author{Paul Cassak\orcidlink{0000-0002-5938-1050}}
\affiliation{Department of Physics and Astronomy, Clemson University, Clemson, SC 29634, USA}

\author{Trevor Bowen\orcidlink{0000-0002-4625-3332}}
\affiliation{Space Sciences Laboratory, University of California Berkeley, CA 94720-7450, USA}

\author{Yi Qi\orcidlink{0000-0002-0959-3450}}
\affiliation{Laboratory for Atmospheric and Space Physics, University of Colorado, Boulder, CO 80303, USA}

\date{\today}

\begin{abstract}
In collisionless plasmas, the particle distributions are often far from local thermodynamic equilibrium, providing a significant source of free energy that leads to complex velocity-space dynamics. In turbulent plasmas, the possibility of a cascade of the free energy in velocity space, analogous to the spatial cascade of fluid and electromagnetic field energies, has been suggested. However, observational evidence of plasma turbulence throughout phase space remains limited. Using measurements from the \textit{Magnetospheric Multiscale} mission in a special campaign in the Earth's magnetosheath, we present the first statistical study of signatures consistent with velocity-space turbulence across a wide range of plasma conditions. We observe two main populations in the ion velocity-space spectrum -- an isotropic $-3/2$ spectrum and an anisotropic $-2$ spectrum organized by the ion $\beta$ and bulk flow. In addition, the spectral power laws in the spatial and velocity domains are uncorrelated, while the velocity-space spectral index is correlated with estimated real-space energy transfer rates. These results reveal how the ion spectral properties are systematically organized by plasma conditions and related to real-space turbulence. This connection has broad implications for turbulence and the energy partition between thermal and non-thermal particles in space and astrophysical plasmas.
\end{abstract}

%  \keywords{Suggested keywords}
\maketitle

\textit{Introduction}---Turbulence is a fundamental energy transfer mechanism in fluids and plasmas \cite{frisch1995a,biskamp2008a,brandenburg2013a,falceta-goncalves+2014a}. Through turbulence, energy injected into the system is transferred to small spatial scales until it is dissipated by collisions or kinetic dynamics \cite{verscharen+2019b,pezzi+2021a}. However, the energy transfer does not necessarily occur only in space. Plasma dynamics is governed by the Vlasov-Boltzmann equation
\begin{equation}
\label{equation:vlasov-boltzmann}
    \frac{\partial f}{\partial t}
    +\vb{v}\cdot\grad f
    +\frac{q}{m}\qty(\vb{E}+\vb{v}\times\vb{B})\cdot\frac{\partial f}{\partial\vb{v}}
    =\qty(\frac{\partial f}{\partial t})_\text{c}
\end{equation}
describing the kinetic evolution of a distribution $f(\vb{r},\vb{v},t)$ of particles with mass $m$ and charge $q$ in a magnetic field $\vb{B}(\vb{r},t)$ and electric field $\vb{E}(\vb{r},t)$. In space and astrophysical plasmas, the collisional term $(\partial f/\partial t)_c$ is often weak, resulting in fluctuations in $f$ far from local thermodynamic equilibrium (LTE) \cite{cassak+2023a}, providing a source of free energy to the plasma \cite{schekochihin+2008a}. As the system evolves, the free energy is expended to produce kinetic phenomena with distinct velocity-space signatures \cite{stix1992a,bowen+2022a,huang+2024b,howes2024a}. Thus, collisionless plasmas can have a significant amount of free energy in $f$, leading to highly complex velocity-space dynamics.

With this insight, an emerging theory proposes the possibility of turbulence in velocity space due to non-LTE fluctuations, expanding our understanding of the energy transfer through phase space \cite{tatsuno+2009b,schekochihin+2016a,servidio+2017a,eyink+2018a,nastac+2024a}. In velocity-space turbulence theory, the complexity of $f(\vb{r},\vb{v},t)$ is usually studied through an expansion into coefficients $\hat{f}(\vb{r},\vb{q},t)$ using Hermite, Laguerre, or Hankel transforms, depending on the system's geometry. Similar to the wavenumber $\vb{k}$ conjugate to position $\vb{r}$ under Fourier transform, $\vb{q}$ is a spectral mode conjugate to the velocity $\vb{v}$. In analogy with the classical Kolmogorov phenomenology \cite{frisch1995a}, this theory predicts a cascade of the free energy in velocity mode $\vb{q}$, along with the cascade of fluid and electromagnetic field energies in spatial mode $\vb{k}$. In this phenomenology, this velocity-space cascade is characterized by a power law of the form $\langle|\hat{f}(\vb{r},\vb{q},t)|^2\rangle_{\vb{r},t}\sim q^\gamma$, where $\langle\cdot\rangle_{\vb{r},t}$ is the spatial and temporal average.

Turbulence in velocity space has been investigated in many numerical studies \cite{parker2015a,hatch+2014a,cerri+2018a,pezzi+2018a,pezzi+2019a,celebre+2023a}. However, due to the lack of high-resolution and 3D measurements of $f$, observational evidence of this phenomenon has been sparse and restricted to case studies \cite{servidio+2017a,wu+2023a,larosa+2025a}. Most notably, \citet{servidio+2017a}, hereafter referred to as S17, studied velocity-space turbulence in the Earth's magnetosheath using 3D ion measurements from the \textit{Magnetospheric Multiscale} (MMS) mission \cite{burch+2016a}. They observed a $\gamma=-3/2$ scaling consistent with the prediction for dominant phase-mixing and electric field terms (\nth{2} and \nth{3} terms on the left-hand side (LHS) of \cref{equation:vlasov-boltzmann}) associated with linear particle motions. A background magnetic field (\nth{4} term) introduces 3D effects to the dynamics through cyclotron motion. A $\gamma=-2$ scaling has been associated with strongly magnetized plasmas, as suggested by hybrid simulations \cite{pezzi+2018a} and recently observed in the solar wind \cite{larosa+2025a}. Other predictions for electrostatic turbulence have also been suggested \cite{schekochihin+2016a,nastac+2024a,adkins+2025a,ewart+2025a,celebre+2026a}.

Results from S17 have sparked much interest in observations of velocity-space turbulence, demonstrating that MMS magnetosheath measurements are highly suitable for studying this phenomenon. However, many questions remain unexplored. The plasma $\beta$, i.e., the ratio between the plasma pressure $P$ and magnetic pressure $B_0^2/2\mu_0$ ($B_0$ is the background magnetic field), determines the relative importance of kinetic and magnetic effects. Thus, $\beta$ is a major control parameter for velocity-space turbulence. However, this phenomenon has not been observed in a wide range of $\beta$. Additionally, in a strong $B_0$, differences between linear and cyclotron motions imply that the velocity-space spectrum should exhibit anisotropy, a feature that has likewise not been extensively observed. Another open question concerns how turbulence in velocity space is related to turbulence in real space.

Here, we provide the first large-scale statistical analysis of the ion velocity-space spectrum using MMS measurements across a wide range of magnetosheath conditions. We determine how the ion spectral scaling and anisotropy are organized by plasma conditions and examine its statistical relation to real-space turbulence.

\textit{Data}---The highest-resolution MMS data in burst mode are manually selected to capture events of interest, e.g., magnetic reconnection, shocks, plasma boundaries. Thus, burst-mode data contain a bias that may not reflect typical plasma conditions. To enable unbiased studies of plasma properties in the magnetosheath, MMS carried out an Unbiased Magnetosheath Campaign, where the magnetosheath was sampled in regular burst-mode intervals after crossing the bow shock and before reaching the magnetopause \cite{svenningsson+2024a,svenningsson+2025a,richard+2025b,pecora+2026a,qi+2026a}. Magnetic field measurements are provided by the Flux Gate Magnetometer (FGM) \cite{russell+2016a}. The ion velocity distribution $f(\vb{v},t)$ and moments are provided by the Fast Plasma Investigation (FPI) \cite{pollock+2016a}. In total, 653 intervals are collected from 37 magnetosheath traversals, which amounts to a sample size of $f$ about 600 times larger than that considered in S17. Our processing of $f$ measurements closely follows S17. In the Supplemental Material \cite{supplemental}, we additionally quantify the systematic errors induced by the complete processing pipeline and establish the reliable spectral range for the present analysis.

\begin{figure}[t]
\includegraphics[width=\columnwidth]{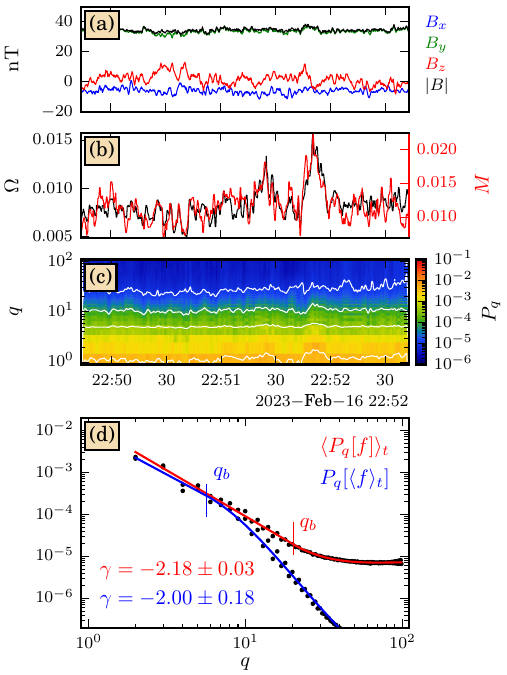}
\caption{An interval in the unbiased magnetosheath campaign. (a) Magnetic field in GSE coordinates. (b) Dimensionless enstrophy $\Omega$ (black) and non-Maxwellianity $M$ (red). (c) Ion Hermite (velocity-space) spectrum. (d) Time-averaged spectrum (red) and spectrum of the time-average VDF (blue). In (c), the white curves are contours of the Hermite spectrum. In (d), the colored curves and printed text are the fit results. $q_b$ is the spectral break in the fit model.}
\label{figure:interval-overview}
\end{figure}

\cref{figure:interval-overview} shows a typical interval in the unbiased campaign. In \cref{figure:interval-overview}(a), there is a steady background magnetic field $\vb{B}_0(t)$ obtained by lowpass filtering $\vb{B}(t)$ at the correlative frequency determined from its autocorrelation \cite{smith+2018a,stawarz+2019a,vo+2026a}. The relative fluctuation amplitude $\delta B/B_0$ is low, where $\delta B\equiv\sqrt{\langle\norm{\vb{B}(t)-\vb{B}_0(t)}^2\rangle_t}$. To quantify the departure of the ion distribution $f(\vb{v},t)$ from LTE, we use the dimensionless enstrophy defined as \cite{pezzi+2021a}
\begin{equation}
\label{equation:enstrophy}
    \Omega(t)\equiv\frac{v_{th,1}(t)v_{th,2}(t)v_{th,\|}(t)}{n_0^2(t)}\int_{\mathbb{R}^3}d^3v\abs{\delta f(\vb{v},t)}^2,
\end{equation}
where ${\delta f(\vb{v},t)\equiv f(\vb{v},t)-f_0(\vb{v},t)}$ and $f_0$ is a 3D Maxwellian distribution constructed from the integrated moments of $f$.  The field-aligned coordinate (FAC) basis is $(\hat{e}_1,\hat{e}_2,\hat{e}_\|)$, where $\hat{e}_\|\propto\vb{B}_0(t)$, $\hat{e}_2\propto-\vb{u}_i(t)\times\vb{B}_0(t)$ is perpendicular to $\hat{e}_\|$ and the ion bulk flow $\vb{u}_i(t)$, and $\hat{e}_1$ completes the triad. $n_0(t)$ is the density, $v_{th,c}(t)=\sqrt{2k_BT_{c}(t)/m}$ is the thermal speed, and $T_c(t)$ is the temperature along component $c$ in FAC. $\Omega(t)$ is related to the entropy-based non-Maxwellianity \cite{kaufmann2009a}
\begin{equation}
\label{equation:kpl-entropy}
    M(t)\equiv-\frac1{1+\ln[2\pi (v_{th}(t)/\Delta v)^2]}\frac{\delta s(t)}{(3/2)n_0(t)},
\end{equation}
with a finite velocity grid size $\Delta v$ \cite{liang+2020c}. $\delta s(t)\equiv s(t)-s_0(t)$, where $s(t)\equiv-\int_{\mathbb{R}^3}d^3v f\ln f$ is the entropy density and $s_0(t)$ is defined similarly with $f_0$. For small $\delta f$, $\delta s(t)\approx-\int_{\mathbb{R}^3}d^3v|\delta f|^2/2f$, which is similar to $-\Omega(t)$. So $M(t)$ and $\Omega(t)$ are expected to correlate, as in \cref{figure:interval-overview}(b).

Similarly to S17, we use the 3D rectangular Hermite transform to calculate $\hat{f}(\vb{q},t)$, but we also explore the fundamental role of anisotropy \cite{supplemental}. We calculate the time-dependent omni-directional Hermite spectrum \cite{servidio+2017a,wu+2023a,larosa+2025a}, defined by $P_q(t)\equiv\sum_{q-1/2<\norm{\vb{q}'}\leq q+1/2}|\hat{f}(\vb{q}',t)|^2$, where $\vb{q}=(n,m,l)$ is the 3D Hermite mode in FAC. By Parseval's theorem, $P_{q\geq1}$ describes how the non-Maxwellianity measure $\Omega$ is distributed in $q$.

In \cref{figure:interval-overview}(c), fluctuations in $P_q(t)$ correlate with those in $\Omega(t)$ and $M(t)$ in \cref{figure:interval-overview}(b). The largest enhancement in $\Omega(t)$ and $M(t)$ at around 22:51:50 corresponds to the rotation in $B_z$ (red) in \cref{figure:interval-overview}(a). These observations suggest that intermittent structures in the magnetic field can displace the ions from LTE \cite{pezzi+2018a}, enhancing the ion velocity-space structure represented by $P_q(t)$.

\cref{figure:interval-overview}(d) shows the time-averaged omni-directional Hermite spectrum $\langle P_q[f]\rangle_t$ (red) from the interval. At low $q$, $\langle P_q[f]\rangle_t$ exhibits a power law below a break at $q_b\approx20$, beyond which instrumental noise and increasing processing errors take effect \cite{supplemental}. To quantify the Hermite index $\gamma$, we perform a fit for $P_q\sim q^{\gamma}$ (colored curve) \cite{supplemental}. The supplemental error analysis shows that the processing response degrades with increasing $q$, but remains reasonably reliable for $q\lesssim q_b$.

Because magnetosheath ions are generally non-Maxwellian, stationary VDF structures can themselves produce broad and power-law-like spectra. To confirm that the observed spectrum is associated with a non-stationary process, \cref{figure:interval-overview}(d) also shows the spectrum of the time-averaged VDF $P_q[\langle f\rangle_t]$ (blue), which represents persistent VDF structures in this interval. The two spectra are comparable at low order, but their divergence beyond $q\approx 6$ suggests that much of the high-order power in $\langle P_q[f]\rangle_t$ contains non-stationary contributions.

The low $\beta$ in this interval and fitted $\gamma\approx-2$ are consistent with the strongly magnetized regime. Additionally, we fit the parallel and perpendicular spectra $P_n\sim n^{\gamma_\perp}$ and $P_l\sim l^{\gamma_\|}$ to calculate the spectral anisotropy $\gamma_\perp/\gamma_\|$. $P_n$ and $P_l$ are calculated by reducing $P_q(t)$ in the other two dimensions \cite{parker+2016a,servidio+2017a,supplemental,pezzi+2018a}. We repeat these calculations across the campaign to determine the statistical properties of $\gamma$ and $\gamma_\perp/\gamma_\|$.

\begin{figure}[t]
\includegraphics[width=\columnwidth]{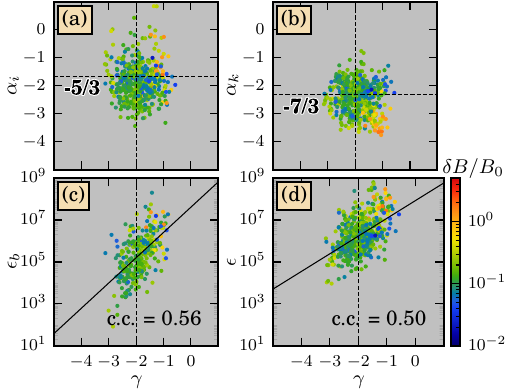}
\caption{Distribution of the Hermite spectral index $\gamma$ and (a) inertial-range and (b) kinetic-range Fourier spectral indices $\alpha_i$ and $\alpha_k$, (c) energy transfer rate in the magnetic field $\epsilon_b$, and (d) total energy transfer rate $\epsilon$. The data are colored based on the relative magnetic field fluctuation $\delta B/B_0$. In (a-b), horizontal dashed lines are relevant power laws for real-space turbulence. In (c-d), the solid black line is the linear regression of the parameters, with printed correlation coefficients.}
\label{figure:fourier-vs-hermite}
\end{figure}

\textit{Relations between real- and velocity-space turbulence}---To characterize real-space turbulence in each interval \cite{supplemental}, we calculate the magnetic field spectrum $P_B(k)\sim k^\alpha$ and determine the Fourier spectral indices $\alpha_i$ and $\alpha_k$ in the inertial range ($k\rho_i\ll 1$; $\rho_i=v_{th}/\omega_{ci}$ is the ion gyroradius and $\omega_{ci}=qB_0/m$ is the ion cyclotron frequency) and (subion) kinetic range ($k\rho_i\gg 1$), respectively. We also calculate the energy transfer rate $\epsilon$ in the inertial range using the Politano-Pouquet law for magnetohydrodynamic (MHD) turbulence \cite{politano1998b,marino2023a}
\begin{equation}
\label{equation:pp-law}
    \langle\delta u_{i,\|}\norm{\delta \vb{u}_i}^2+\delta u_{i,\|}\norm{\delta\vb{b}}^2-2\delta b_\|\qty(\delta \vb{u}_i\cdot\delta\vb{b})\rangle_t=-\frac43\epsilon L,
\end{equation}
where ${\vb{b}\equiv\vb{B}/\sqrt{\mu_0n_0m}}$, ${\delta \bm{\psi}\equiv\bm{\psi}(t+\tau)-\bm{\psi}(t)}$ for quantity $\bm{\psi}(t)$, and ${L\equiv u_i\tau\sim k^{-1}}$. For more details about these calculations, see also \citet{doan+2026a}.

\begin{figure*}
\includegraphics[width=2\columnwidth]{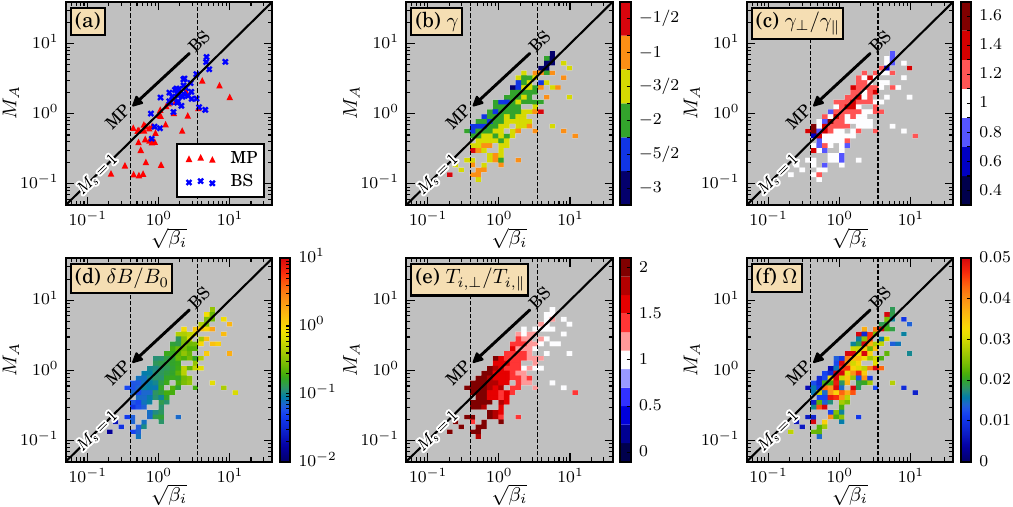}
\caption{Hermite spectral properties as a function of $\sqrt{\beta_i}$ and Alfvén Mach number $M_A$. (a) Magnetopause (MP) and bow shock (BS) intervals. The arrow indicates the direction of traversal. (b) Omni-directional Hermite spectral index $\gamma$. (c) Hermite spectral anisotropy $\gamma_\perp/\gamma_\|$. (d) Magnetic fluctuation amplitude $\delta B/B_0$. (e) Ion temperature anisotropy $T_{i,\perp}/T_{i,\|}$. (f) Enstrophy $\Omega$. Solid black lines indicate the acoustic Mach number $M_s=1$. Dotted vertical lines denote the statistically significant region.}
\label{figure:hermite-statistics} 
\end{figure*}

In \cref{figure:fourier-vs-hermite}, we show distributions of the Hermite spectral index $\gamma$ and Fourier-space parameters (a) $\alpha_i$, (b) $\alpha_k$, (c) the energy transfer rate in the magnetic field $\epsilon_b$ (\nth{2} term on the LHS of \cref{equation:pp-law}), and (d) the total energy transfer rate $\epsilon$. In \cref{figure:fourier-vs-hermite}(a-b), the exponent $\gamma$ averages around $-2$ (vertical dashed line), given the presence of a strong $\vb{B}_0$ in most intervals. The range of $\gamma$ is wide mainly due to the physical variability in plasma conditions in the dataset. Conditions for the asymptotic predictions for $\gamma$ are rarely satisfied. Similarly, the exponents $\alpha_i$ and $\alpha_k$ exhibit wide ranges, but their averages are consistent with the $-5/3$ Kolmogorov scaling \cite{frisch1995a} and $-7/3$ Hall MHD scaling (horizontal dashed lines) \cite{alexandrova+2008a}, respectively. Notably, there is no correlation between Fourier and Hermite indices. This result is expected since Fourier and Hermite transforms operate on independent parts of phase space.

On the contrary, in \cref{figure:fourier-vs-hermite}(c-d), there is a moderate correlation between the Hermite index $\gamma$ and the energy transfer rates. The colors ($\delta B/B_0$) indicate that $\gamma$ tends to be closer to $-3/2$ for high field fluctuation amplitude. These correlations suggest that stronger real-space turbulence is associated with a shallower Hermite spectrum, revealing a statistical connection between real-space energy transfer and velocity-space turbulence, consistent with previous studies \cite{pezzi+2018a,sorriso-valvo+2019a,celebre+2023a}. A extended analysis of fluctuations in both the magnetic field and ion distribution in joint Fourier-Hermite domain will better reveal the characteristics of the energy transfer in phase space. However, such a study is out of scope of the present work.

\textit{Hermite spectral properties}---A major problem in analyzing spatial evolution in the unbiased campaign is that conditions in both the upstream solar wind and the magnetosheath change during a traversal from the bow shock (BS) to the magnetopause (MP) \cite{pecora+2026a}. However, plasma conditions in the first and last interval in each traversal are always well-categorized as those of the BS and MP, respectively. \cref{figure:hermite-statistics}(a) maps the endpoints using the local ion plasma beta $\beta_i$ and Alfvén Mach number $M_A=u_i/V_A$, where $V_A=B_0/\sqrt{\mu_0n_0m}$ is the Alfvén speed. $\sqrt{\beta_i}$ is the relative amplitude between the phase-mixing and magnetic field Hermite time scales \cite{servidio+2017a} in \cref{equation:vlasov-boltzmann}, associated with $\gamma=-3/2$ and $\gamma=-2$ asymptotic power laws. The slope in \cref{figure:hermite-statistics} is the acoustic Mach number $M_s=M_A/\sqrt{\beta_i}=u_i/v_{th}$ characterizing the ion flow. High-Mach magnetosheath jets \cite{plaschke+2018a,kramer+2024a} that could be formed by foreshock transients \cite{raptis+2022a,raptis+2025a,raptis+2026a} can generate waves, reconnection, bow waves, etc, resulting in high non-Maxwellianity \cite{liu+2020b,roberts+2025a}. Thus, the ion flow $u_i$ can be another control parameter for velocity-space turbulence. The distinct BS and MP clusters establish a qualitative statistical spatial trend indicated by the arrow. Most traversals within the dotted region follow this evolution, despite the large variations between the endpoints.

Following the same format as \cref{figure:hermite-statistics}(a), panels (b) and (c) present the omni-directional Hermite spectral index $\gamma$ and the Hermite anisotropy $\gamma_\perp/\gamma_\|$. Panel (d) displays the fluctuation amplitude $\delta B/B_0$, while panel (e) shows the ion temperature anisotropy $T_{i,\perp}/T_{i,\|}$. Finally, panel (f) reports the enstrophy $\Omega$. The fluctuation amplitude in (d) decreases toward the MP, a behavior expected downstream of the BS. In panel (e), the increasing strength of the background geomagnetic field compresses the ions in the direction parallel to the field, leading to a progressively more anisotropic distribution (see also \cite{richard+2025b}).

Crucially, \cref{figure:hermite-statistics}(f) shows that $\Omega$ is non-uniformly distributed and does not spatially correlate with the fluctuation amplitude in \cref{figure:hermite-statistics}(d). Velocity-space turbulence is suggested to be a local phenomenon by previous observations and simulations \cite{pezzi+2018a,wu+2023a}, which explains the non-uniformity of $\Omega$.

In \cref{figure:hermite-statistics}(b-c), we observe two main populations in the Hermite spectral properties. A $\gamma=-2$ magnetic field dominated population in \cref{figure:hermite-statistics}(b) (green) exhibits high anisotropy in \cref{figure:hermite-statistics}(c) (red). A $\gamma=-3/2$ kinetic pressure dominated population (yellow) exhibits low anisotropy (white). The anisotropy is associated with 3D effects introduced by $B_0$. Anisotropic spectra tend to extend preferentially in the direction parallel to $\vb{B}_0$ \cite{pezzi+2018a}. Further, we observe that the transition between these populations is systematically associated with the ion $\beta_i$ and bulk flow $u_i$. At fixed $M_A$, increasing $\beta_i$ marks a transition from the magnetic regime to the kinetic regime. However, the dependence on the ion flow $u_i$ is more enigmatic. The magnetic regime tends to be distributed along $M_s=1$ (black line), while the kinetic regime is mainly subsonic ($M_s<1$).

\textit{Discussion}---In this Letter, we observe two main populations in the ion velocity-space spectrum that are systematically organized by the ion $\beta$ and bulk flow. The spectral properties [\cref{figure:hermite-statistics}(b-c)] are consistent with predictions from velocity-space turbulence theory \cite{servidio+2017a}. The observed spectra are associated with non-stationary processes [\cref{figure:interval-overview}(d)]. Further diagnostics in the Supplemental Material establish that the observed spectral properties are also characteristic of high-enstrophy ion distributions \cite{supplemental}. These results support the interpretation that our observations are consistent with velocity-space turbulence as the guiding theoretical framework.

The bulk flow dependence remains largely unexplored in theories and simulations, and may have broad implications for collisionless shock dynamics. Magnetosheath jets, known to generate waves, reconnection, shock fronts, and turbulence, have also been observed at Mars \cite{gunell+2023a} and Jupiter \cite{zhou+2024a}. From this perspective, our results provide an observational foundation for investigating velocity-space turbulence and particle acceleration in planetary magnetospheres and interplanetary shocks \cite{trotta+2021a,trotta+2022a}.

The Hermite (velocity-space) spectral properties are uncorrelated with Fourier (real-space) spectral properties [\cref{figure:fourier-vs-hermite}(a-b)], global background conditions [\cref{figure:hermite-statistics}(d-e)], and the enstrophy [\cref{figure:hermite-statistics}(f)]. However, shallower Hermite spectra are correlated with higher real-space energy transfer rates [\cref{figure:fourier-vs-hermite}(c-d)]. These results suggest that real- and velocity-space turbulence signatures need not be correlated to remain energetically related. Velocity-space properties do not appear to evolve spatially across the magnetosheath. This observation motivates future studies of localized structures, including reconnection and ion beams associated with real-space turbulence \cite{sorriso-valvo+2019a}, in the context of velocity-space turbulence.

Other theoretically proposed spectral indices \cite{schekochihin+2016a,nastac+2024a,adkins+2025a,ewart+2025a,celebre+2026a} are less prominent in the campaign. A minor $\gamma=-5/2$ population appears mainly for $M_s>1$ \cite{supplemental}, while no distinct $\gamma=-1/2$ population is observed. These scalings arise under more restrictive electrostatic conditions and may not be prominent on a statistical level in the generally electromagnetic magnetosheath turbulence. However, our results do not rule out the presence of these regimes in individual magnetosheath intervals.

The impact of this study is that it provides statistical evidence consistent with velocity-space turbulence across broad plasma conditions in the Earth's magnetosheath. The observed Hermite index and anisotropy reveal how non-Maxwellian VDF features are structured in velocity space and related to real-space energy transfer, highlighting the importance of kinetic and non-LTE physics in plasma turbulence \cite{cassak+2023a}. Beyond their consistency with previous theoretical and numerical studies, our results reveal a novel dependence on the ion bulk flow and provide observational foundations for future studies on velocity-space turbulence. These results contribute to the long-lasting problem of the energy partition between thermal and non-thermal particles in diverse contexts, including particle acceleration \cite{trotta+2021a,juno+2023a} and downstream thermalization \cite{ghavamian+2013a,raymond+2023a} at heliospheric and astrophysical shocks.

\begin{acknowledgments}
This research was supported by the International Space Science Institute (ISSI) in Bern, through the ISSI International Team project \#23-588 (Unveiling Energy Conversion and Dissipation in Non-Equilibrium Space Plasmas). This work utilized the Alpine high performance computing resource at the University of Colorado Boulder. Alpine is jointly funded by the University of Colorado Boulder, the University of Colorado Anschutz, Colorado State University, and the National Science Foundation (award 2201538).

TV, RE, AC, and YQ are supported by NASA-MMS Contract NNG04EB99C. TV is also supported by NASA-FINESST Grant 80NSSC23K1630. AC acknowledges support from NASA Grants 80NSSC21K0454, 80NSSC22K0688, and 80NSSC24K0172. OP and AL are supported by the FIS2 Starting Grant FIS-2023-00246 “PhAse-sPAce cOmplexity in turbulent nearly-reversible plasmas (PAPAO)” (CUP B53C24009610001) funded by the Italian Ministry of University and Research. LSV was supported by the Swedish National Space Agency (SNSA) Research Grant N. 2025-00229. LSV and OP were supported by the Swedish Research Council (VR) Research Grant N. 2022-03352. LSV and SS are supported by Agenzia Spaziale Italiana (ASI) project: “Radial Evolution of large- and KInetic-scale Processes in the Expanding Solar wind (REKIPES)" - C83C25000900005. PC is supported by NASA Grants 80NSSC23K0409 and 80NSSC24K0173, and NSF Grant PHY-2607428. TB is supported by NASA HSR grant 80NSSC24K0272. MMS Level 2 FIELDS and FPI data products are used in this study. The authors thank the entire MMS team for their work on the mission, and Savvas Raptis and Domenico Trotta for helpful discussions.
\end{acknowledgments}

The data behind the figures in this article are openly available at \cite{vo+2026}, embargo periods may apply. The data that support the findings of this article are available upon reasonable request from the authors.

\nocite{*}
\bibliography{ref}

@misc{supplemental,
  note = "See Supplemental Material at
    [url] for details about the error analysis of the calculated Hermite spectra, fit model, and analyses of the spectral anisotropy, non-stationarity, and robustness of the reported statistical results, which includes Refs. [14-15, 34, 36, 44-46, 68, 72-73, 75-76, 78, 81-82].",
    year = 2026,
}

@dataset{vo+2026,
  author       = {Vo, Tien},
  title        = {Data behind figures for "Statistical Signatures of
                   Velocity-space Turbulence in the Earth's
                   Magnetosheath"
                  },
  month        = oct,
  year         = 2026,
  publisher    = {Zenodo},
  doi          = {10.5281/zenodo.23821840},
  url          = {https://doi.org/10.5281/zenodo.23821840},
}

@article{adkins+2025a,
  title = {Turbulent {{Heating}} in {{Collisionless Low-beta Plasmas}}: {{Imbalance}}, {{Landau Damping}}, and {{Electron}}--{{Ion Energy Partition}}},
  shorttitle = {Turbulent {{Heating}} in {{Collisionless Low-beta Plasmas}}},
  author = {Adkins, Toby and Meyrand, Romain and Squire, Jonathan},
  year = 2025,
  month = sep,
  journal = {The Astrophysical Journal},
  volume = {990},
  number = {2},
  pages = {138},
  publisher = {The American Astronomical Society},
  issn = {0004-637X},
  doi = {10.3847/1538-4357/ade9b0},
  urldate = {2026-09-09},
  langid = {english}
}

@article{alexandrova+2008a,
  title = {Small-{{Scale Energy Cascade}} of the {{Solar Wind Turbulence}}},
  author = {Alexandrova, O. and Carbone, V. and Veltri, P. and Sorriso-Valvo, L.},
  year = 2008,
  month = feb,
  journal = {The Astrophysical Journal},
  volume = {674},
  number = {2},
  pages = {1153--1157},
  issn = {0004-637X, 1538-4357},
  doi = {10.1086/524056},
  urldate = {2023-09-11},
  langid = {english}
}

@book{biskamp2008a,
  title = {Magnetohydrodynamic {{Turbulence}}},
  author = {Biskamp, Dieter},
  year = 2008,
  publisher = {Cambridge University Press}
}

@article{bowen+2022a,
  title = {In {{Situ Signature}} of {{Cyclotron Resonant Heating}} in the {{Solar Wind}}},
  author = {Bowen, Trevor A. and Chandran, Benjamin D. G. and Squire, Jonathan and Bale, Stuart D. and Duan, Die and Klein, Kristopher G. and Larson, Davin and Mallet, Alfred and McManus, Michael D. and Meyrand, Romain and Verniero, Jaye L. and Woodham, Lloyd D.},
  year = 2022,
  month = oct,
  journal = {Physical Review Letters},
  volume = {129},
  number = {16},
  pages = {165101},
  publisher = {American Physical Society},
  doi = {10.1103/PhysRevLett.129.165101},
  urldate = {2026-02-20}
}

@article{brandenburg2013a,
  title = {Astrophysical {{Hydromagnetic Turbulence}}},
  author = {Brandenburg, A. and Lazarian, A.},
  year = 2013,
  month = oct,
  journal = {Space Science Reviews},
  volume = {178},
  number = {2},
  pages = {163--200},
  issn = {1572-9672},
  doi = {10.1007/s11214-013-0009-3},
  urldate = {2025-07-23},
  langid = {english}
}

@article{burch+2016a,
  title = {Magnetospheric {{Multiscale Overview}} and {{Science Objectives}}},
  author = {Burch, J. L. and Moore, T. E. and Torbert, R. B. and Giles, B. L.},
  year = 2016,
  month = mar,
  journal = {Space Science Reviews},
  volume = {199},
  number = {1-4},
  pages = {5--21},
  issn = {0038-6308, 1572-9672},
  doi = {10.1007/s11214-015-0164-9},
  urldate = {2023-09-11},
  langid = {english}
}

@article{cassak+2023a,
  title = {Quantifying {{Energy Conversion}} in {{Higher-Order Phase Space Density Moments}} in {{Plasmas}}},
  author = {Cassak, Paul A. and Barbhuiya, M. Hasan and Liang, Haoming and Argall, Matthew R.},
  year = 2023,
  month = feb,
  journal = {Physical Review Letters},
  volume = {130},
  number = {8},
  pages = {085201},
  issn = {0031-9007, 1079-7114},
  doi = {10.1103/PhysRevLett.130.085201},
  urldate = {2026-02-02},
  langid = {english}
}

@article{celebre+2023a,
  title = {Phase Space Dynamics of Unmagnetized Plasmas: {{Collisionless}} and Collisional Regimes},
  shorttitle = {Phase Space Dynamics of Unmagnetized Plasmas},
  author = {Celebre, G. and Servidio, S. and Valentini, F.},
  year = 2023,
  month = sep,
  journal = {Physics of Plasmas},
  volume = {30},
  number = {9},
  pages = {092304},
  issn = {1070-664X, 1089-7674},
  doi = {10.1063/5.0160549},
  urldate = {2025-11-11},
  langid = {english}
}

@misc{celebre+2026a,
  title = {Plasma Turbulence Driven by Wave-Hole Interaction},
  author = {Celebre, Gabriele and Imbrogno, Mario and Servidio, Sergio and Valentini, Francesco},
  year = 2026,
  month = jun,
  number = {arXiv:2606.22098},
  eprint = {2606.22098},
  primaryclass = {physics.plasm-ph},
  publisher = {arXiv},
  doi = {10.48550/arXiv.2606.22098},
  urldate = {2026-09-09},
  archiveprefix = {arXiv}
}

@article{cerri+2018a,
  title = {Dual {{Phase-space Cascades}} in {{3D Hybrid-Vlasov}}--{{Maxwell Turbulence}}},
  author = {Cerri, S. S. and Kunz, M. W. and Califano, F.},
  year = 2018,
  month = mar,
  journal = {The Astrophysical Journal Letters},
  volume = {856},
  number = {1},
  pages = {L13},
  publisher = {The American Astronomical Society},
  issn = {2041-8205},
  doi = {10.3847/2041-8213/aab557},
  urldate = {2026-02-20},
  langid = {english}
}

@article{doan+2026a,
  title = {Statistical Correlations between Turbulent Structures and Dissipation Proxies in Near-{{Earth}} Space: The {{MMS}} 2023-2024 Unbiased Magnetosheath Campaign},
  author = {Doan, Viet and {Sorriso-Valvo}, Luca and Raynal, Pierre and {Montagud-Camps}, Victor and Settino, Adriana and Tamburrini, Abiam and Svenningsson, Ida and Chasapis, Alexandros and Vo, Tien and Yordanova, Emiliya and Richard, Louis and Pecora, Francesco and V{\"o}r{\"o}s, Zoltan and Servidio, Sergio and Pezzi, Oreste},
  year = 2026,
  journal = {in preparation}
}

@article{ewart+2025a,
  title = {Relaxation to Universal Non-{{Maxwellian}} Equilibria in a Collisionless Plasma},
  author = {Ewart, Robert J. and Nastac, Michael L. and Bilbao, Pablo J. and Silva, Thales and Silva, Lu{\'i}s O. and Schekochihin, Alexander A.},
  year = 2025,
  month = apr,
  journal = {Proceedings of the National Academy of Sciences},
  volume = {122},
  number = {17},
  pages = {e2417813122},
  issn = {0027-8424, 1091-6490},
  doi = {10.1073/pnas.2417813122},
  urldate = {2026-09-09},
  langid = {english}
}

@article{eyink+2018a,
  title = {Cascades and {{Dissipative Anomalies}} in {{Nearly Collisionless Plasma Turbulence}}},
  author = {Eyink, Gregory L.},
  year = 2018,
  month = nov,
  journal = {Physical Review X},
  volume = {8},
  number = {4},
  pages = {041020},
  publisher = {American Physical Society},
  doi = {10.1103/PhysRevX.8.041020},
  urldate = {2026-05-04}
}

@article{falceta-goncalves+2014a,
  title = {Turbulence in the Interstellar Medium},
  author = {{Falceta-Gon{\c c}alves}, D. and Kowal, G. and Falgarone, E. and Chian, A. C.-L.},
  year = 2014,
  month = may,
  journal = {Nonlinear Processes in Geophysics},
  volume = {21},
  number = {3},
  pages = {587--604},
  publisher = {Copernicus GmbH},
  issn = {1023-5809},
  doi = {10.5194/npg-21-587-2014},
  urldate = {2025-07-30},
  langid = {english}
}

@book{frisch1995a,
  title = {Turbulence: {{The Legacy}} of {{A}}. {{N}}. {{Kolmogorov}}},
  author = {Frisch, Uriel},
  year = 1995,
  publisher = {Cambridge University Press}
}

@article{ghavamian+2013a,
  title = {Electron-{{Ion Temperature Equilibration}} in {{Collisionless Shocks}}: {{The Supernova Remnant-Solar Wind Connection}}},
  shorttitle = {Electron-{{Ion Temperature Equilibration}} in {{Collisionless Shocks}}},
  author = {Ghavamian, Parviz and Schwartz, Steven J. and Mitchell, Jeremy and Masters, Adam and Laming, J. Martin},
  year = 2013,
  month = oct,
  journal = {Space Science Reviews},
  volume = {178},
  number = {2},
  pages = {633--663},
  issn = {1572-9672},
  doi = {10.1007/s11214-013-9999-0},
  urldate = {2026-05-06},
  langid = {english}
}

@article{gunell+2023a,
  title = {Magnetosheath Jets at {{Mars}}},
  author = {Gunell, Herbert and Hamrin, Maria and {Nesbit-{\"O}stman}, Sara and Kr{\"a}mer, Eva and Nilsson, Hans},
  year = 2023,
  month = jun,
  journal = {Science Advances},
  volume = {9},
  number = {22},
  pages = {eadg5703},
  issn = {2375-2548},
  doi = {10.1126/sciadv.adg5703},
  urldate = {2026-05-06},
  langid = {english}
}

@article{hatch+2014a,
  title = {Phase Space Scales of Free Energy Dissipation in Gradient-Driven Gyrokinetic Turbulence},
  author = {Hatch, D. R. and Jenko, F. and Bratanov, V. and Navarro, A. Ba{\~n}{\'o}n},
  year = 2014,
  month = aug,
  journal = {Journal of Plasma Physics},
  volume = {80},
  number = {4},
  pages = {531--551},
  issn = {0022-3778, 1469-7807},
  doi = {10.1017/S0022377814000154},
  urldate = {2026-01-20},
  langid = {english}
}

@article{howes2024a,
  title = {The Fundamental Parameters of Astrophysical Plasma Turbulence and Its Dissipation: Non-Relativistic Limit},
  shorttitle = {The Fundamental Parameters of Astrophysical Plasma Turbulence and Its Dissipation},
  author = {Howes, Gregory G.},
  year = 2024,
  month = oct,
  journal = {Journal of Plasma Physics},
  volume = {90},
  number = {5},
  pages = {905900504},
  issn = {0022-3778, 1469-7807},
  doi = {10.1017/S0022377824001090},
  urldate = {2025-04-09},
  langid = {english}
}

@article{huang+2024b,
  title = {The Velocity-Space Signature of Transit-Time Damping},
  author = {Huang, Rui and Howes, Gregory G. and McCubbin, Andrew J.},
  year = 2024,
  month = aug,
  journal = {Journal of Plasma Physics},
  volume = {90},
  number = {4},
  pages = {535900401},
  issn = {0022-3778, 1469-7807},
  doi = {10.1017/S0022377824000667},
  urldate = {2026-03-26},
  langid = {english}
}

@article{juno+2023a,
  title = {Phase-Space {{Energization}} of {{Ions}} in {{Oblique Shocks}}},
  author = {Juno, James and Brown, Collin R. and Howes, Gregory G. and Haggerty, Colby C. and TenBarge, Jason M. and Wilson III, Lynn B. and Caprioli, Damiano and Klein, Kristopher G.},
  year = 2023,
  month = feb,
  journal = {The Astrophysical Journal},
  volume = {944},
  number = {1},
  pages = {15},
  publisher = {The American Astronomical Society},
  issn = {0004-637X},
  doi = {10.3847/1538-4357/acaf53},
  urldate = {2026-05-06},
  langid = {english}
}

@article{kaufmann2009a,
  title = {Boltzmann {{H}} Function and Entropy in the Plasma Sheet},
  author = {Kaufmann, Richard L. and Paterson, William R.},
  year = 2009,
  journal = {Journal of Geophysical Research: Space Physics},
  volume = {114},
  number = {A9},
  issn = {2156-2202},
  doi = {10.1029/2008JA014030},
  urldate = {2025-12-18},
  copyright = {Copyright 2009 by the American Geophysical Union.},
  langid = {english}
}

@article{kramer+2024a,
  title = {Jets {{Downstream}} of {{Collisionless Shocks}}: {{Recent Discoveries}} and {{Challenges}}},
  shorttitle = {Jets {{Downstream}} of {{Collisionless Shocks}}},
  author = {Kr{\"a}mer, Eva and Koller, Florian and Suni, Jonas and LaMoury, Adrian T. and P{\"o}ppelwerth, Adrian and Glebe, Georg and {Mohammed-Amin}, Tara and Raptis, Savvas and Vuorinen, Laura and Weiss, Stefan and Xirogiannopoulou, Niki and Archer, Martin and {Blanco-Cano}, X{\'o}chitl and Gunell, Herbert and Hietala, Heli and Karlsson, Tomas and Plaschke, Ferdinand and Preisser, Luis and Roberts, Owen and Simon~Wedlund, Cyril and Temmer, Manuela and V{\"o}r{\"o}s, Zolt{\'a}n},
  year = 2024,
  month = dec,
  journal = {Space Science Reviews},
  volume = {221},
  number = {1},
  pages = {4},
  issn = {1572-9672},
  doi = {10.1007/s11214-024-01129-3},
  urldate = {2026-05-06},
  langid = {english}
}

@article{larosa+2025a,
  title = {Velocity-Space {{Turbulent Cascade}} in the {{Near-Sun Solar Wind}}: {{First Insights}} from the {{Parker Solar Probe Mission}}},
  shorttitle = {Velocity-Space {{Turbulent Cascade}} in the {{Near-Sun Solar Wind}}},
  author = {Larosa, A. and Pezzi, O. and Bowen, T. and {Sorriso-Valvo}, L. and Sioulas, N. and Pucci, F. and Trotta, D. and Verniero, J. L. and Livi, R. and Das, S. Bharati and Chasapis, A. and Perrone, D. and Valentini, F. and Servidio, S.},
  year = 2025,
  month = dec,
  journal = {The Astrophysical Journal Letters},
  volume = {995},
  number = {2},
  pages = {L69},
  publisher = {The American Astronomical Society},
  issn = {2041-8205},
  doi = {10.3847/2041-8213/ae2869},
  urldate = {2026-01-08},
  langid = {english}
}

@article{liang+2020c,
  title = {Kinetic Entropy-Based Measures of Distribution Function Non-{{Maxwellianity}}: Theory and Simulations},
  shorttitle = {Kinetic Entropy-Based Measures of Distribution Function Non-{{Maxwellianity}}},
  author = {Liang, Haoming and Barbhuiya, M. Hasan and Cassak, P. A. and Pezzi, O. and Servidio, S. and Valentini, F. and Zank, G. P.},
  year = 2020,
  month = oct,
  journal = {Journal of Plasma Physics},
  volume = {86},
  number = {5},
  pages = {825860502},
  issn = {0022-3778, 1469-7807},
  doi = {10.1017/S0022377820001270},
  urldate = {2025-12-18},
  langid = {english}
}

@article{liu+2020b,
  title = {Statistical {{Study}} of {{Magnetosheath Jet-Driven Bow Waves}}},
  author = {Liu, Terry Z. and Hietala, Heli and Angelopoulos, Vassilis and Omelchenko, Yuri and Vainio, Rami and Plaschke, Ferdinand},
  year = 2020,
  journal = {Journal of Geophysical Research: Space Physics},
  volume = {125},
  number = {7},
  pages = {e2019JA027710},
  issn = {2169-9402},
  doi = {10.1029/2019JA027710},
  urldate = {2026-05-07},
  langid = {english}
}

@article{marino2023a,
  title = {Scaling Laws for the Energy Transfer in Space Plasma Turbulence},
  author = {Marino, Raffaele and {Sorriso-Valvo}, Luca},
  year = 2023,
  month = mar,
  journal = {Physics Reports},
  volume = {1006},
  pages = {1--144},
  issn = {03701573},
  doi = {10.1016/j.physrep.2022.12.001},
  urldate = {2025-03-07},
  langid = {english}
}

@article{nastac+2024a,
  title = {Phase-Space Entropy Cascade and Irreversibility of Stochastic Heating in Nearly Collisionless Plasma Turbulence},
  author = {Nastac, Michael L. and Ewart, Robert J. and Sengupta, Wrick and Schekochihin, Alexander A. and Barnes, Michael and Dorland, William D.},
  year = 2024,
  month = jun,
  journal = {Physical Review E},
  volume = {109},
  number = {6},
  pages = {065210},
  publisher = {American Physical Society},
  doi = {10.1103/PhysRevE.109.065210},
  urldate = {2026-02-20}
}

@article{parker+2016a,
  title = {Suppression of Phase Mixing in Drift-Kinetic Plasma Turbulence},
  author = {Parker, J. T. and Highcock, E. G. and Schekochihin, A. A. and Dellar, P. J.},
  year = 2016,
  month = jul,
  journal = {Physics of Plasmas},
  volume = {23},
  number = {7},
  pages = {070703},
  issn = {1070-664X},
  doi = {10.1063/1.4958954},
  urldate = {2026-01-20}
}

@article{parker2015a,
  title = {Fourier-{{Hermite}} Spectral Representation for the {{Vlasov-Poisson}} System in the Weakly Collisional Limit},
  author = {Parker, Joseph T. and Dellar, Paul J.},
  year = 2015,
  journal = {Journal of Plasma Physics},
  volume = {81},
  number = {305810203},
  doi = {10.1017/S0022377814001287},
  langid = {english}
}

@article{pecora+2026a,
  title = {General {{Properties}} of the {{Magnetosheath}}: {{The MMS Unbiased Campaign}}},
  shorttitle = {General {{Properties}} of the {{Magnetosheath}}},
  author = {Pecora, Francesco and Chhiber, Rohit and Roy, Sohom and Chasapis, Alexandros and Vo, Tien and {Sorriso-Valvo}, Luca and {Montagud-Camps}, Victor and Wang, Jiaming and Matthaeus, William H.},
  year = 2026,
  journal = {Journal of Geophysical Research: Space Physics},
  volume = {131},
  number = {3},
  pages = {e2025JA034919},
  issn = {2169-9402},
  doi = {10.1029/2025JA034919},
  urldate = {2026-04-08},
  copyright = {\copyright{} 2026. American Geophysical Union. All Rights Reserved.},
  langid = {english}
}

@article{pezzi+2018a,
  title = {Velocity-Space Cascade in Magnetized Plasmas: {{Numerical}} Simulations},
  shorttitle = {Velocity-Space Cascade in Magnetized Plasmas},
  author = {Pezzi, O. and Servidio, S. and Perrone, D. and Valentini, F. and {Sorriso-Valvo}, L. and Greco, A. and Matthaeus, W. H. and Veltri, P.},
  year = 2018,
  month = jun,
  journal = {Physics of Plasmas},
  volume = {25},
  number = {6},
  pages = {060704},
  issn = {1070-664X, 1089-7674},
  doi = {10.1063/1.5027685},
  urldate = {2024-09-11},
  langid = {english}
}

@article{pezzi+2019a,
  title = {Proton--{{Proton Collisions}} in the {{Turbulent Solar Wind}}: {{Hybrid Boltzmann}}--{{Maxwell Simulations}}},
  shorttitle = {Proton--{{Proton Collisions}} in the {{Turbulent Solar Wind}}},
  author = {Pezzi, O. and Perrone, D. and Servidio, S. and Valentini, F. and {Sorriso-Valvo}, L. and Veltri, P.},
  year = 2019,
  month = dec,
  journal = {The Astrophysical Journal},
  volume = {887},
  number = {2},
  pages = {208},
  publisher = {The American Astronomical Society},
  issn = {0004-637X},
  doi = {10.3847/1538-4357/ab5285},
  urldate = {2026-02-20},
  langid = {english}
}

@article{pezzi+2021a,
  title = {Dissipation Measures in Weakly Collisional Plasmas},
  author = {Pezzi, O and Liang, H and Juno, J L and Cassak, P A and V{\'a}sconez, C L and {Sorriso-Valvo}, L and Perrone, D and Servidio, S and Roytershteyn, V and TenBarge, J M and Matthaeus, W H},
  year = 2021,
  month = aug,
  journal = {Monthly Notices of the Royal Astronomical Society},
  volume = {505},
  number = {4},
  pages = {4857--4873},
  issn = {0035-8711},
  doi = {10.1093/mnras/stab1516},
  urldate = {2025-12-18}
}

@article{plaschke+2018a,
  title = {Jets {{Downstream}} of {{Collisionless Shocks}}},
  author = {Plaschke, Ferdinand and Hietala, Heli and Archer, Martin and {Blanco-Cano}, X{\'o}chitl and Kajdi{\v c}, Primo{\v z} and Karlsson, Tomas and Lee, Sun Hee and Omidi, Nojan and Palmroth, Minna and Roytershteyn, Vadim and Schmid, Daniel and Sergeev, Victor and Sibeck, David},
  year = 2018,
  month = jun,
  journal = {Space Science Reviews},
  volume = {214},
  number = {5},
  pages = {81},
  issn = {1572-9672},
  doi = {10.1007/s11214-018-0516-3},
  urldate = {2026-05-06},
  langid = {english}
}

@article{politano1998b,
  title = {Von {{K\'arm\'an}}--{{Howarth}} Equation for Magnetohydrodynamics and Its Consequences on Third-Order Longitudinal Structure and Correlation Functions},
  author = {Politano, H. and Pouquet, A.},
  year = 1998,
  month = jan,
  journal = {Physical Review E},
  volume = {57},
  number = {1},
  pages = {R21-R24},
  publisher = {American Physical Society (APS)},
  issn = {1063-651X, 1095-3787},
  doi = {10.1103/physreve.57.r21},
  urldate = {2025-07-09},
  copyright = {http://link.aps.org/licenses/aps-default-license},
  langid = {english}
}

@article{pollock+2016a,
  title = {Fast {{Plasma Investigation}} for {{Magnetospheric Multiscale}}},
  author = {Pollock, C. and Moore, T. and Jacques, A. and Burch, J. and Gliese, U. and Saito, Y. and Omoto, T. and Avanov, L. and Barrie, A. and Coffey, V. and Dorelli, J. and Gershman, D. and Giles, B. and Rosnack, T. and Salo, C. and Yokota, S. and Adrian, M. and Aoustin, C. and Auletti, C. and Aung, S. and Bigio, V. and Cao, N. and Chandler, M. and Chornay, D. and Christian, K. and Clark, G. and Collinson, G. and Corris, T. and De~Los~Santos, A. and Devlin, R. and Diaz, T. and Dickerson, T. and Dickson, C. and Diekmann, A. and Diggs, F. and Duncan, C. and {Figueroa-Vinas}, A. and Firman, C. and Freeman, M. and Galassi, N. and Garcia, K. and Goodhart, G. and Guererro, D. and Hageman, J. and Hanley, J. and Hemminger, E. and Holland, M. and Hutchins, M. and James, T. and Jones, W. and Kreisler, S. and Kujawski, J. and Lavu, V. and Lobell, J. and LeCompte, E. and Lukemire, A. and MacDonald, E. and Mariano, A. and Mukai, T. and Narayanan, K. and Nguyan, Q. and Onizuka, M. and Paterson, W. and Persyn, S. and Piepgrass, B. and Cheney, F. and Rager, A. and Raghuram, T. and Ramil, A. and Reichenthal, L. and Rodriguez, H. and Rouzaud, J. and Rucker, A. and Saito, Y. and Samara, M. and Sauvaud, J.-A. and Schuster, D. and Shappirio, M. and Shelton, K. and Sher, D. and Smith, D. and Smith, K. and Smith, S. and Steinfeld, D. and Szymkiewicz, R. and Tanimoto, K. and Taylor, J. and Tucker, C. and Tull, K. and Uhl, A. and Vloet, J. and Walpole, P. and Weidner, S. and White, D. and Winkert, G. and Yeh, P.-S. and Zeuch, M.},
  year = 2016,
  month = mar,
  journal = {Space Science Reviews},
  volume = {199},
  number = {1-4},
  pages = {331--406},
  issn = {0038-6308, 1572-9672},
  doi = {10.1007/s11214-016-0245-4},
  urldate = {2023-09-11},
  langid = {english}
}

@article{qi+2026a,
  title = {The {{Evolution}} of {{Parallel Electric Field Across}} the {{Earth}}'s {{Magnetosheath}}},
  author = {Qi, Y. and Ergun, R. E. and Vo, T. and Chasapis, A. and Schwartz, S. J. and Wilder, V.},
  year = 2026,
  journal = {Geophysical Research Letters},
  volume = {53},
  number = {19},
  pages = {e2026GL123482},
  issn = {1944-8007},
  doi = {10.1029/2026GL123482},
  urldate = {2026-10-26},
  langid = {english}
}

@article{raptis+2022a,
  title = {Downstream High-Speed Plasma Jet Generation as a Direct Consequence of Shock Reformation},
  author = {Raptis, Savvas and Karlsson, Tomas and Vaivads, Andris and Pollock, Craig and Plaschke, Ferdinand and Johlander, Andreas and Trollvik, Henriette and Lindqvist, Per-Arne},
  year = 2022,
  month = feb,
  journal = {Nature Communications},
  volume = {13},
  number = {1},
  pages = {598},
  publisher = {Nature Publishing Group},
  issn = {2041-1723},
  doi = {10.1038/s41467-022-28110-4},
  urldate = {2026-05-07},
  copyright = {2022 The Author(s)},
  langid = {english}
}

@article{raptis+2025a,
  title = {Multimission {{Observations}} of {{Relativistic Electrons}} and {{High-speed Jets Linked}} to {{Shock-generated Transients}}},
  author = {Raptis, Savvas and Lindberg, Martin and Liu, Terry Z. and Turner, Drew L. and Lalti, Ahmad and Zhou, Yufei and Kajdi{\v c}, Primo{\v z} and Kouloumvakos, Athanasios and Sibeck, David G. and Vuorinen, Laura and Michael, Adam and Shumko, Mykhaylo and Osmane, Adnane and Kr{\"a}mer, Eva and Turc, Lucile and Karlsson, Tomas and Katsavrias, Christos and Wilson, Lynn B. and Madanian, Hadi and {Blanco-Cano}, X{\'o}chitl and Cohen, Ian J. and Escoubet, C. Philippe},
  year = 2025,
  month = feb,
  journal = {The Astrophysical Journal Letters},
  volume = {981},
  number = {1},
  pages = {L10},
  publisher = {The American Astronomical Society},
  issn = {2041-8205},
  doi = {10.3847/2041-8213/adb154},
  urldate = {2026-05-07},
  langid = {english}
}

@article{raptis+2026a,
  title = {Compressive {{Structures}} in the {{Foreshock}} of {{Collisionless Shocks}}},
  author = {Raptis, Savvas and Trotta, Domenico and Turner, Drew L. and {Blanco-Cano}, X{\'o}chitl and Hietala, Heli and Karlsson, Tomas and Jebaraj, Immanuel Christopher and Vasko, Ivan Y. and Osmane, Adnane and Takahashi, Kazue and Lario, David and Wilson, Lynn B. and Howes, Gregory G. and {Wimmer-Schweingruber}, Robert F.},
  year = 2026,
  month = mar,
  journal = {The Astrophysical Journal Letters},
  volume = {1000},
  number = {2},
  pages = {L55},
  publisher = {The American Astronomical Society},
  issn = {2041-8205},
  doi = {10.3847/2041-8213/ae53e5},
  urldate = {2026-05-07},
  langid = {english}
}

@article{raymond+2023a,
  title = {Electron--{{Ion Temperature Ratio}} in {{Astrophysical Shocks}}},
  author = {Raymond, John C. and Ghavamian, Parviz and Bohdan, Artem and Ryu, Dongsu and Niemiec, Jacek and Sironi, Lorenzo and Tran, Aaron and Amato, Elena and Hoshino, Masahiro and Pohl, Martin and Amano, Takanobu and Fiuza, Frederico},
  year = 2023,
  month = may,
  journal = {The Astrophysical Journal},
  volume = {949},
  number = {2},
  pages = {50},
  publisher = {The American Astronomical Society},
  issn = {0004-637X},
  doi = {10.3847/1538-4357/acc528},
  urldate = {2026-05-06},
  langid = {english}
}

@article{richard+2025b,
  title = {Non-{{Maxwellianity}} of Ion Velocity Distributions in the {{Earth}}'s Magnetosheath},
  author = {Richard, Louis and Servidio, Sergio and Svenningsson, Ida and Artemyev, Anton V. and Klein, Kristopher G. and Yordanova, Emiliya and Chasapis, Alexandros and Pezzi, Oreste and Khotyaintsev, Yuri V.},
  year = 2025,
  month = nov,
  journal = {Physical Review E},
  volume = {112},
  number = {5},
  pages = {L053201},
  issn = {2470-0045, 2470-0053},
  doi = {10.1103/7p1x-84y9},
  urldate = {2025-12-03},
  langid = {english}
}

@article{roberts+2025a,
  title = {Energy {{Conversion}} and {{Exchange}} in a {{Magnetosheath Jet}}},
  author = {Roberts, O. W. and V{\"o}r{\"o}s, Z. and Settino, A. and Koller, F. and Raptis, S. and Temmer, M. and Wedlund, C. Simon and Li, X. and Nakamura, R.},
  year = 2025,
  journal = {Journal of Geophysical Research: Space Physics},
  volume = {130},
  number = {11},
  pages = {e2025JA034414},
  issn = {2169-9402},
  doi = {10.1029/2025JA034414},
  urldate = {2026-05-06},
  copyright = {\copyright{} 2025. The Author(s).},
  langid = {english}
}

@article{russell+2016a,
  title = {The {{Magnetospheric Multiscale Magnetometers}}},
  author = {Russell, C. T. and Anderson, B. J. and Baumjohann, W. and Bromund, K. R. and Dearborn, D. and Fischer, D. and Le, G. and Leinweber, H. K. and Leneman, D. and Magnes, W. and Means, J. D. and Moldwin, M. B. and Nakamura, R. and Pierce, D. and Plaschke, F. and Rowe, K. M. and Slavin, J. A. and Strangeway, R. J. and Torbert, R. and Hagen, C. and Jernej, I. and Valavanoglou, A. and Richter, I.},
  year = 2016,
  month = mar,
  journal = {Space Science Reviews},
  volume = {199},
  number = {1-4},
  pages = {189--256},
  issn = {0038-6308, 1572-9672},
  doi = {10.1007/s11214-014-0057-3},
  urldate = {2023-09-11},
  langid = {english}
}

@article{schekochihin+2008a,
  title = {Gyrokinetic Turbulence: A Nonlinear Route to Dissipation through Phase Space},
  shorttitle = {Gyrokinetic Turbulence},
  author = {Schekochihin, A A and Cowley, S C and Dorland, W and Hammett, G W and Howes, G G and Plunk, G G and Quataert, E and Tatsuno, T},
  year = 2008,
  month = nov,
  journal = {Plasma Physics and Controlled Fusion},
  volume = {50},
  number = {12},
  pages = {124024},
  issn = {0741-3335},
  doi = {10.1088/0741-3335/50/12/124024},
  urldate = {2026-02-01},
  langid = {english}
}

@article{schekochihin+2016a,
  title = {Phase Mixing versus Nonlinear Advection in Drift-Kinetic Plasma Turbulence},
  author = {Schekochihin, A. A. and Parker, J. T. and Highcock, E. G. and Dellar, P. J. and Dorland, W. and Hammett, G. W.},
  year = 2016,
  month = apr,
  journal = {Journal of Plasma Physics},
  volume = {82},
  number = {2},
  pages = {905820212},
  issn = {0022-3778, 1469-7807},
  doi = {10.1017/S0022377816000374},
  urldate = {2025-11-14},
  copyright = {https://www.cambridge.org/core/terms},
  langid = {english}
}

@article{servidio+2017a,
  title = {Magnetospheric {{Multiscale Observation}} of {{Plasma Velocity-Space Cascade}}: {{Hermite Representation}} and {{Theory}}},
  shorttitle = {Magnetospheric {{Multiscale Observation}} of {{Plasma Velocity-Space Cascade}}},
  author = {Servidio, S. and Chasapis, A. and Matthaeus, W. H. and Perrone, D. and Valentini, F. and Parashar, T. N. and Veltri, P. and Gershman, D. and Russell, C. T. and Giles, B. and Fuselier, S. A. and Phan, T. D. and Burch, J.},
  year = 2017,
  month = nov,
  journal = {Physical Review Letters},
  volume = {119},
  number = {20},
  pages = {205101},
  issn = {0031-9007, 1079-7114},
  doi = {10.1103/PhysRevLett.119.205101},
  urldate = {2025-05-20},
  copyright = {https://link.aps.org/licenses/aps-default-license},
  langid = {english}
}

@article{smith+2018a,
  title = {Correlation {{Scales}} of the {{Turbulent Cascade}} at 1 Au},
  author = {Smith, Charles W. and Vasquez, Bernard J. and Coburn, Jesse T. and Forman, Miriam A. and Stawarz, Julia E.},
  year = 2018,
  month = apr,
  journal = {The Astrophysical Journal},
  volume = {858},
  number = {1},
  pages = {21},
  publisher = {The American Astronomical Society},
  issn = {0004-637X},
  doi = {10.3847/1538-4357/aabb00},
  urldate = {2025-07-03},
  langid = {english}
}

@article{sorriso-valvo+2019a,
  title = {Turbulence-{{Driven Ion Beams}} in the {{Magnetospheric Kelvin-Helmholtz Instability}}},
  author = {{Sorriso-Valvo}, Luca and Catapano, Filomena and Retin{\`o}, Alessandro and Le Contel, Olivier and Perrone, Denise and Roberts, Owen W. and Coburn, Jesse T. and Panebianco, Vincenzo and Valentini, Francesco and Perri, Silvia and Greco, Antonella and Malara, Francesco and Carbone, Vincenzo and Veltri, Pierluigi and Pezzi, Oreste and Fraternale, Federico and Di Mare, Francesca and Marino, Raffaele and Giles, Barbara and Moore, Thomas E. and Russell, Christopher T. and Torbert, Roy B. and Burch, Jim L. and Khotyaintsev, Yuri V.},
  year = 2019,
  month = jan,
  journal = {Physical Review Letters},
  volume = {122},
  number = {3},
  pages = {035102},
  publisher = {American Physical Society},
  doi = {10.1103/PhysRevLett.122.035102},
  urldate = {2026-04-08}
}

@article{stawarz+2019a,
  title = {Properties of the {{Turbulence Associated}} with {{Electron-only Magnetic Reconnection}} in {{Earth}}'s {{Magnetosheath}}},
  author = {Stawarz, J. E. and Eastwood, J. P. and Phan, T. D. and Gingell, I. L. and Shay, M. A. and Burch, J. L. and Ergun, R. E. and Giles, B. L. and Gershman, D. J. and Contel, O. Le and Lindqvist, P.-A. and Russell, C. T. and Strangeway, R. J. and Torbert, R. B. and Argall, M. R. and Fischer, D. and Magnes, W. and Franci, L.},
  year = 2019,
  month = jun,
  journal = {The Astrophysical Journal Letters},
  volume = {877},
  number = {2},
  pages = {L37},
  issn = {2041-8205, 2041-8213},
  doi = {10.3847/2041-8213/ab21c8},
  urldate = {2025-02-17},
  langid = {english}
}

@book{stix1992a,
  title = {Waves in {{Plasmas}}},
  author = {Stix, Thomas Howard},
  year = 1992,
  publisher = {American Institute of Physics}
}

@article{tatsuno+2009b,
  title = {Nonlinear Phase Mixing and Phase-Space Cascade of Entropy in Gyrokinetic Plasma Turbulence},
  author = {Tatsuno, T. and Dorland, W. and Schekochihin, A A and Plunk, G G and Barnes, M. and Cowley, S. C. and Howes, G. G.},
  year = 2009,
  journal = {Physical Review Letters},
  volume = {103},
  number = {1},
  publisher = {American Physical Society}
}

\end{document}